\documentclass[a4paper]{jpconf}
\usepackage{graphicx}
\usepackage{amsmath}
\usepackage{epsf}
\usepackage{epsfig}
\usepackage{amssymb}
\usepackage{epstopdf}
\usepackage{latexsym}
\usepackage{subfigure}
\usepackage{setspace}
\begin{document}
\title{Constraints on $D$ Dimensional Warped Spaces.}

\author{Paul R.~Archer}

\address{\em  
$^a$Department of Physics \& Astronomy, University of Sussex, Brighton
BN1 9QH, UK}

\ead{p.archer@sussex.ac.uk}

\begin{abstract}
In order to investigate the phenomenological implications of allowing gauge fields to propagate in warped spaces of more than five dimensions, we consider a toy model of a space warped by the presence of a anisotropic bulk cosmological constant. After solving the Einstein equation, three classes of solutions are found, those in which the additional ($D>5$) dimensions are growing, shrinking or remaining constant. It is found that gauge fields propagating in these spaces have a significantly different Kaluza Klein (KK) mass spectrum and couplings from that of the Randall and Sundrum model. This leads to a greatly reduced lower bound on the KK scale, arising from electroweak constraints, for spaces growing towards the IR brane.   
\end{abstract}

\section{Introduction}
The Randall and Sundrum (RS) model \cite{Randall:1999ee} has proved a popular extension to the standard model for a number of reasons. Firstly it offers a non supersymmetric resolution to the gauge hierarchy problem and secondly it provides a natural model with which to describe flavour physics \cite{Grossman:1999ra, Gherghetta:2000qt, Huber:2000ie}. However allowing the standard model gauge fields to propagate in the bulk gives rise to corrections to the electroweak observables (EWO's) at tree level, which can naively force the lower bound on the Kaluza Klein (KK) mass scale to be $M_{KK}\gtrsim 11$ TeV, corresponding to a mass of the first gauge boson excitation of about $27$ TeV \cite{Csaki:2002gy}. This bound can be reduced to $M_{KK}\gtrsim 2-3$ TeV  by allowing the fermions to propagate in the bulk and localising them towards the UV brane and at the same time introducing either large brane kinetic terms \cite{Carena:2002dz} or a bulk  $SU(2)_R\times SU(2)_L\times U(1)$ gauge symmetry \cite{Agashe:2003zs}.   

In spite of this, before the RS model can be considered a full resolution to the hierarchy problem it requires a UV completion. There has been considerable work on providing a UV completion via string theory, particularly with regard to the AdS/CFT correspondence, for example \cite{ArkaniHamed:2000ds, PerezVictoria:2001pa, Reece:2010xj}, but this work necessarily requires models of more than five dimensions. Although the RS model is often seen as the low energy effective theory of $\rm{AdS}_5\times S^5$, it is quite possible that the $D>5$ additional dimensions are also warped \cite{Bao:2005ni, Klebanov:2000hb}.

With this in mind we first look generally at what really determines the size of these EW corrections, before taking a `bottom up' approach to extending the RS model to more than 5D, by considering a space with a anisotropic bulk cosmological constant.

\section{General Warped Extra Dimensions}
In this section we consider a $4+1+\delta$ dimensional spaced warped with respect to (w.r.t) a single `preferred' direction $r$, described by
\begin{equation}
\label{Metric}
ds^2=a^2(r)\eta_{\mu\nu}dx^\mu dx^\nu-b^2(r)dr^2-c^2(r)d\Omega_\delta^2,
\end{equation}
where $\eta_{\mu\nu}=\mbox{diag}(+---)$ is the 4D Minkowski metric and $d\Omega_\delta^2=\gamma_{ij}d\phi^id\phi^j$, with $i,j$ running from $1\dots \delta$. As in the RS model the space is bounded by two branes at $r= r_{\rm{ir}}$ and $r= r_{\rm{uv}}$.  
\subsection{Resolving the Hierarchy Problem}
From a phenomenological perspective, one of the central ideas of extra dimensional models is that the 4D bare parameters of a given theory are not fundamental but have been scaled down from their higher dimensional values. Hence to resolve the hierarchy problem it is required that the dimensionful parameters of the higher dimensional theory be all at the same order of magnitude or in particular that the fundamental Higgs mass ($m_{\rm{fund}}^{{\rm{Higgs}}}$) be similar to the fundamental Planck mass ($M_{\rm{fund}}$). In computing the 4D effective parameters there are essentially two effects at work. Firstly the bare parameters are scaled by the volume of the extra dimension. So for example the fundamental Planck mass is scaled by
\begin{equation}
\label{PlankMass}
M_{\rm{P}}^2\sim\int d^{\delta+1}x\; a^2bc^\delta \sqrt{\gamma}\;M_{\rm{Fund}}^{\delta +3}.
\end{equation} 
This is the mechanism used in the ADD model \cite{ArkaniHamed:1998rs, Antoniadis:1998ig}. Secondly fields localised throughout the space would have masses suppressed by gravitational red shifting or warping. So the Higgs, localised on the IR brane, would have a 4D effective mass 
\begin{equation}
\label{HiggsMass}
m_{\rm{4D}}^2=a^2(r_{\rm{ir}})m_{\rm{fund}}^2.
\end{equation}    
In the RS model the volume effects are $\mathcal{O}(1)$ and hence a warp factor of $a(r_{\rm{ir}})^{-1}\equiv\Omega\sim 10^{15}$ is required in order to resolve the hierarchy problem \cite{Randall:1999ee}. Although clearly these mechanisms are not mutually exclusive and the fundamental scale of nature could be neither the Planck scale nor the EW scale but some intermediate scale.  

\subsection{Electroweak Corrections}
Likewise when one computes the EW corrections it is found that they are largely determined by the extent to which the 4D gauge couplings have been scaled. To be more precise, after making the usual KK decomposition
\begin{equation} \label{KKexp}
A_\mu=\sum_{n}A_\mu^{(n)}(x^\mu)f_n(r)\Theta_n(\phi_1,\dots,\phi_\delta) \mbox{  such that }\int d^{1+\delta}x\; bc^\delta\sqrt{\gamma}f_nf_m\Theta_n\Theta_m=\delta_{nm},
\end{equation}
where the profiles are given by
\begin{equation} \label{KKprof}
f_n^{\prime\prime}+\frac{(a^2b^{-1}c^\delta)^{\prime}}{(a^2b^{-1}c^\delta)}f_n^{\prime}-\frac{b^2}{c^2}\alpha_nf_n+\frac{b^2}{a^2}m_n^2f_n=0,\hspace{0.5cm}-\frac{1}{\sqrt{\gamma}}\partial_{\phi_i}(\sqrt{\gamma}\gamma^{ij}\partial_{\phi_j}\Theta_n)=\alpha_n\Theta_n,
\end{equation}
and $^{\prime}$ denotes derivative w.r.t $r$. Then the tree level corrections to the EWO's are determined by three parameters, the coupling of the KK gauge fields to the Higgs relative to that of the zero mode,
\begin{equation}
F_n\equiv\frac{f_{n}(r_{\rm{ir}})}{f_{0}(r_{\rm{ir}})}\quad \mbox{or}\quad F_n\equiv\frac{f_{n}(r_{\rm{ir}})\Theta_n(\phi_{\rm{ir}})}{f_{0}(r_{\rm{ir}})\Theta_0(\phi_{\rm{ir}})},
\end{equation}
the coupling of the KK gauge fields to the fermions relative to that of the zero mode
\begin{equation}
F_\psi^{(n)}\equiv\frac{f_\psi^{(0,n,0)}}{f_\psi^{(0,0,0)}}\quad\mbox{where}\quad f_\psi^{(l,n,m)}\equiv \int\, d^{\delta+1}x\, ba^3c^\delta\sqrt{\gamma}\,f_{L}^{(l)}\Theta_L^{(l)}\,f_{n}\Theta_n\,f_{L}^{(m)}\Theta_L^{(m)}.
\end{equation}
and the mass of the KK gauge fields $m_n^2$. Here $f_{L}^{(n)}$ and $\Theta_L^{(n)}$ are the KK profiles of the fermions, analogous to (\ref{KKexp}). The two definitions of the gauge-Higgs coupling are related to whether the Higgs is localised on a 3 brane or just w.r.t $r$ as explained in \cite{Archer:2010hh}.

Often the EW corrections are parametrised in terms of the Peskin-Takeuchi parameters  \cite{Peskin:1991sw}, however the approach taken in \cite{Archer:2010hh} was to fix input parameters by comparison with  $\hat{\alpha}(M_Z)^{-1}$, $G_f$ and $\hat{M}_Z$ and then compute corrections to eight EWO's. In practice the resulting constraints from the two methods do not differ significantly. It is found that typically the tightest constraints come from the weak mixing angle ($s_Z^2$), which when there is a bulk $SU(2)\times U(1)$ gauge symmetry is given by
\begin{equation}
s_Z^2\approx s_p^2\left (1-\frac{c_p^2}{c_p^2-s_p^2}\sum_{n=1}\left [\frac{m_z^2F_n^2}{m_n^2}-\frac{m_w^2\left(F_n-F_\psi^{(n)}\right )^2}{m_n^2}\right ]+\mathcal{O}(m_n^{-4})\right ).
\end{equation}
While if there is a bulk $SU(2)_R \times SU(2)_L \times U(1)$ gauge symmetry it is given by
\begin{equation}
\label{SZcust}
s_Z^2\approx s_p^2\left (1+\frac{c_p^2}{c_p^2-s_p^2}\sum_{n=1}\left [\frac{m_w^2F_\psi^{(n)\,2}}{m_n^2}-\frac{2m_w^2F_nF_\psi^{(n)}}{m_n^2}+s^{\prime\, 2}m_w^2\left (\frac{\tilde{F}_n^2}{\tilde{m}_n^2}-\frac{F_n^2}{m_n^2}\right )+\mathcal{O}(m_n^{-4})\right ]\right ),
\end{equation}
where $s^\prime$ is the mixing angle associated with the breaking of $SU(2)_R\times U(1)\rightarrow U(1)$, $\tilde{F}_n^2$ and $\tilde{m}_n^2$ are the equivalent couplings and masses for the $SU(2)_R$ gauge fields,  $ s_p^2=\frac{1}{2}\left (1-\sqrt{1-\frac{4\pi\alpha}{\sqrt{2}G_f\hat{M}_Z^2}}\right )$ and $c_p^2=1-s_p^2$. Note that when $F_n\sim \tilde{F}_n$ and $m_n\sim \tilde{m}_n$ then the fourth term of (\ref{SZcust}) cancels and the constraints become linearly dependent on $F_n$. This term is essentially the remnant of the Peskin-Takeuchi T parameter.

\subsection{Reducing the Gauge-Higgs Coupling in 5D}
Typically $F_\psi$ can always be reduced by localising the fermions towards the UV brane away from the KK gauge fields and hence the lower bound on the KK mass scale is dominated by the value of $F_n$. For the RS model (with $\Omega=10^{15}$) $F_n\approx 8.3$, while for universal extra dimensions $F_n=\sqrt{2}$. One can then ask the question, whether a space can be found in which the KK gauge fields effectively decouple from the Higgs? In five dimensions the gauge Higgs coupling is given by
\begin{equation}
F_n=\frac{\sqrt{\int b\,dr}\, f_n(r_{\rm{ir}})}{\sqrt{\int b f_n^2\,dr}}.
\end{equation}
However it can be shown that the warp factor is related to the gauge profile by
\begin{equation}
\Omega^2=\frac{b^2(r_{\rm{uv}})f_n(r_{\rm{uv}})f_n^{\prime\prime}(r_{\rm{ir}})}{b^2(r_{\rm{ir}})f_n(r_{\rm{ir}})f_n^{\prime\prime}(r_{\rm{uv}})}.
\end{equation}
Hence if one requires a large warp factor to resolve the hierarchy problem, then one would need a very unusually shaped profile, in order to reduce the gauge-Higgs coupling.  So in agreement with \cite{Delgado:2007ne}, one can conclude that five dimensional spaces that resolve the hierarchy problem via warping, will typically have large EW corrections.   

\begin{figure}
\begin{center}
\includegraphics[width=3.9in]{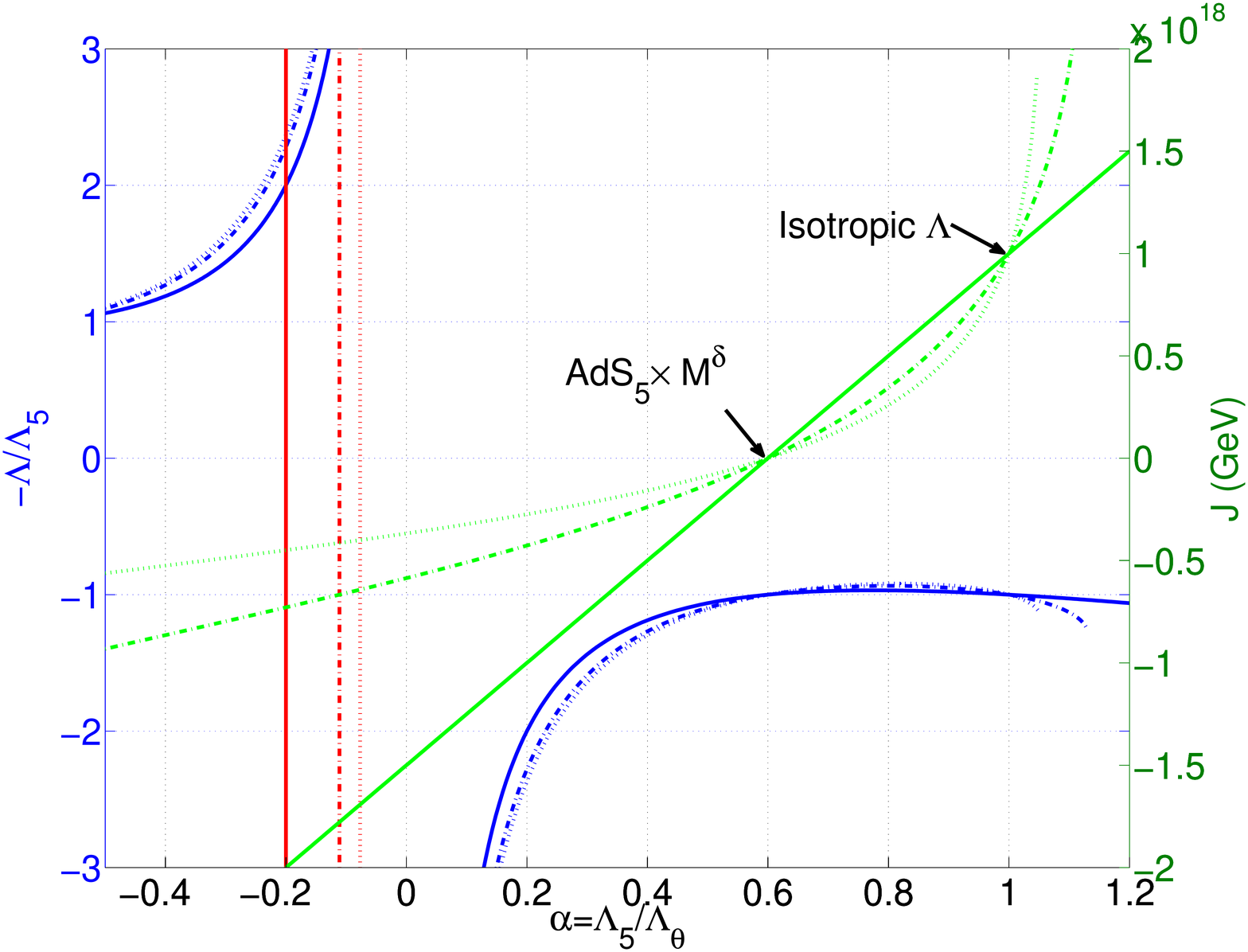}
\caption{\footnotesize The solutions to the Einstein equations for 6D (solid lines), 8D(dash-dot lines) and 10D (dot-dot lines). $J$ is plotted in green and $-\frac{\Lambda}{\Lambda_5}$ in blue. The red lines correspond to $2k+\delta J=0$ and hence the spaces to the left of them will have exponentially suppressed fundamental Planck masses. Here we have fixed $\Omega\equiv e^{kR}=10^{15}$ and $M_{KK}\equiv \frac{k}{\Omega}=1$ TeV.\normalsize }
\label{JLamb}
\end{center}
\end{figure}

\section{A D dimensional Extension of the RS Model}
We now consider a $4+1+\delta$ dimensional space, bounded by two 3 branes, described by  
\begin{eqnarray}
S=\int d^{5+\delta}x\sqrt{G}\left [\Lambda-\frac{1}{2}M^{3+\delta}R+\mathcal{L}_{\rm{bulk}}\right]+\int d^4x\sqrt{g_{\rm{ir}}}\left [\mathcal{L}_{\rm{ir}}+V_{\rm{ir}}\right ] \nonumber \\
 +\int d^4x\sqrt{g_{\rm{uv}}}\left [\mathcal{L}_{\rm{uv}}+V_{\rm{uv}}\right ]\label{action},
\end{eqnarray}   
where the co-ordinates run over $(x_\mu,r,\theta_1\dots\theta_\delta)$, $r\in [0,R]$ and $\theta_i\in [0,R_\theta]$. We allow for an anisotropic bulk cosmological constant of the form $\Lambda=\rm{diag}\left (\Lambda \eta_{\mu\nu },\Lambda_5,\Lambda_\theta,\dots,\Lambda_\theta\right )$. We also introduce the parameter $\alpha \equiv \frac{\Lambda_5}{\Lambda_\theta}$. It can then be shown that the Einstein equations admit solutions (although admittedly not unique solutions) of the form
\begin{equation}   
ds^2=e^{-2kr}\eta_{\mu\nu}dx^\mu dx^\nu-dr^2-\sum_{i=1}^{\delta}e^{-2Jr}d\theta_i^2.
\end{equation} 
These solutions are plotted in figure \ref{JLamb}. There are basically three classes of solution. When $\alpha >\frac{3}{5}$, $J>0$ and the warping of the $\delta$ additional dimensions are shrinking towards the IR brane. When $\alpha=\frac{3}{5}$ the $\delta$ additional dimensions are not warped. This scenario has been studied in \cite{Davoudiasl:2002wz, Davoudiasl:2008qm}. Finally when  $\alpha <\frac{3}{5}$, $J<0$ and the warping of the $\delta$ additional dimensions are growing towards the IR brane. Spaces to the left of the vertical (red) lines in figure \ref{JLamb}, would have volume suppressed Planck masses according to (\ref{PlankMass}) and hence risk introducing an additional hierarchy between the AdS curvature and the Planck mass.

\begin{figure}
\begin{center}
\begin{tabular}{cc}
\includegraphics[width=2.8in]{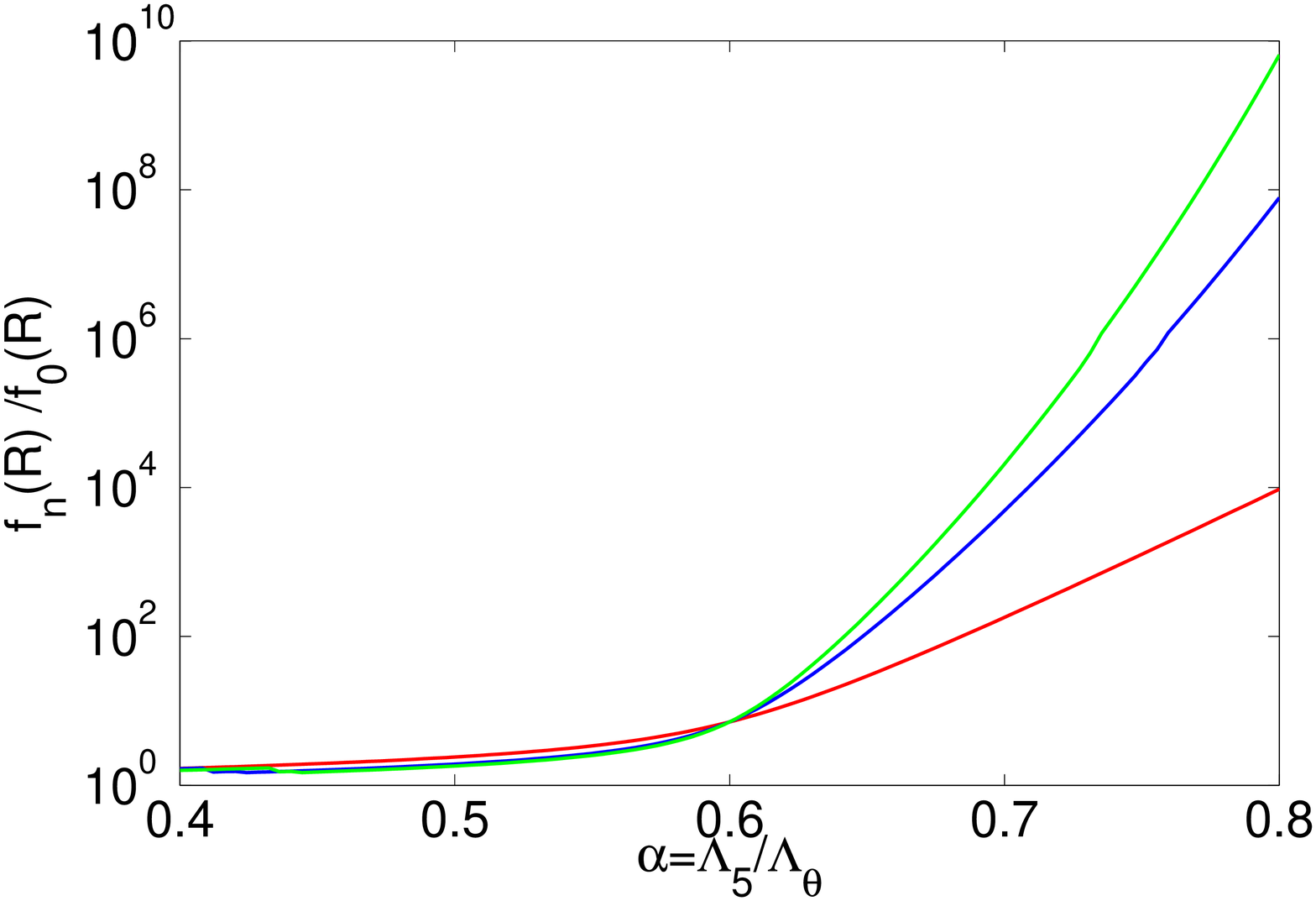}&
\includegraphics[width=2.8in]{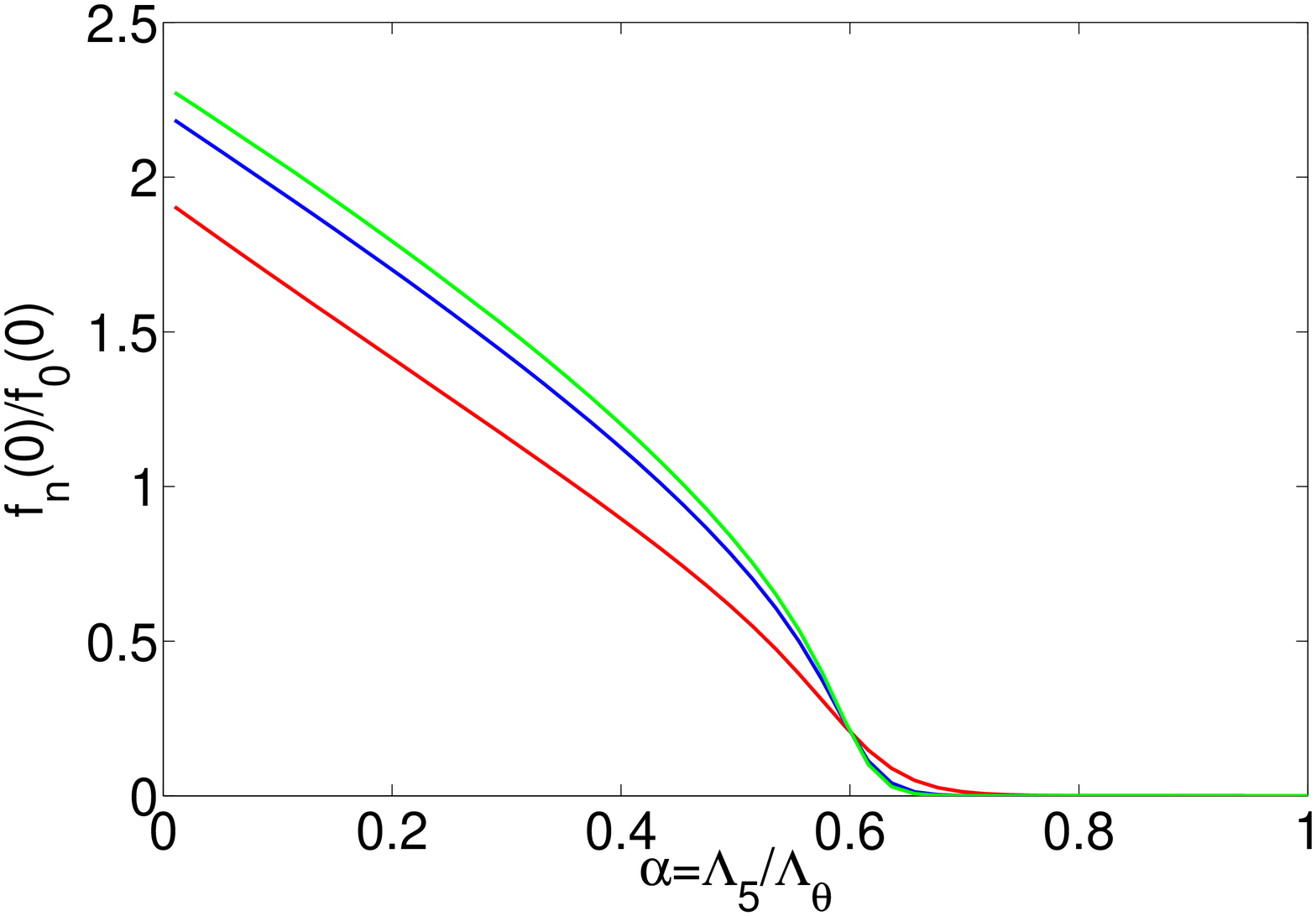}\\
\end{tabular}
\caption{\footnotesize The relative coupling of the $l_i=0$ KK gauge modes with a IR localised Higgs / fermion (on the left) and a UV localised fermion (on the right) in 6D (red), 8D (blue) and 10D (green). Here $\Omega\equiv e^{kR}=10^{15}$ and $M_{KK}\equiv\frac{k}{\Omega}=1$TeV.\normalsize  }
\label{couplingfig }
\end{center}
\end{figure}
Moving now to consider gauge fields propagating in these spaces. The gauge KK profiles are given by (\ref{KKprof}),
\begin{equation}
f_n^{\prime\prime}-(2k+\delta J)f_n^{\prime}-\sum_{i=1}^\delta e^{2Jr}\frac{l_i^2}{R_\theta^2}f_n+e^{2kr}m_n^2f_n=0,\label{gaugeeqn}
\end{equation} 
where $\partial_i^2\Theta_n=-\frac{l_i^2}{R_\theta^2}\Theta_n$. When $l_i\neq 0$, (\ref{gaugeeqn}) has no analytical solutions, although to get a feel for the solutions one can make the substitution $x^2=\left (e^{2kr}m_n^2-e^{2Jr}\sum_{i=1}^\delta\frac{l_i^2}{R_\theta^2}\right )/\gamma^2$ such that $x^{\prime}=\gamma x$.
So now when $\gamma \approx$ constant, (\ref{gaugeeqn}) can be solved to give
\begin{equation}
f(x)\approx Nx^{\frac{1}{2}\frac{2k+\delta J}{\gamma}}\left (\mathbf{J}_{-\frac{1}{2}\frac{2k+\delta J}{\gamma}}(x)+\beta \mathbf{Y}_{-\frac{1}{2}\frac{2k+\delta J}{\gamma}}(x) \right ).
\end{equation}
Typically this approximation is valid in the IR region where the gauge fields are localised and also (\ref{gaugeeqn}) can be solved numerically to check these results \cite{Archer:2010}. As in the RS model, the KK mass eigenvalues are then determined by the zeros of the Bessel functions, hence
\begin{equation}\label{MNapprox}
m_n\sim X_n\frac{\sqrt{\gamma^2+\frac{e^{2JR}}{R_\theta^2}\sum_i^{\delta}l_i^2}}{e^{kR}},
\end{equation}  
where $X_n \sim \mathcal{O}(1)$. Now returning to the three classes of solutions found in the previous section, when $J>0$ (i.e. $\alpha>\frac{3}{5}$) then $\frac{e^{JR}}{R_\theta}\gg\gamma\sim k$ and the $l_i\neq 0$ KK modes would gain masses far larger than $M_{KK}=k/\Omega$ and essentially decouple from the low energy theory. On the other hand if $J\sim 0$ then $\frac{e^{JR}}{R_\theta}\sim\gamma$ and the $l_i\neq 0$ KK modes would have masses of $\mathcal{O}(M_{KK})$. However of potential interest to LHC phenomenologists, is the case when $J<0$, $\frac{e^{JR}}{R_\theta}\ll\gamma$ and a hyperfine splitting in the KK mass spectrum would be introduced.

\begin{figure}
\begin{center}
\begin{tabular}{cc}
\includegraphics[width=2.8in]{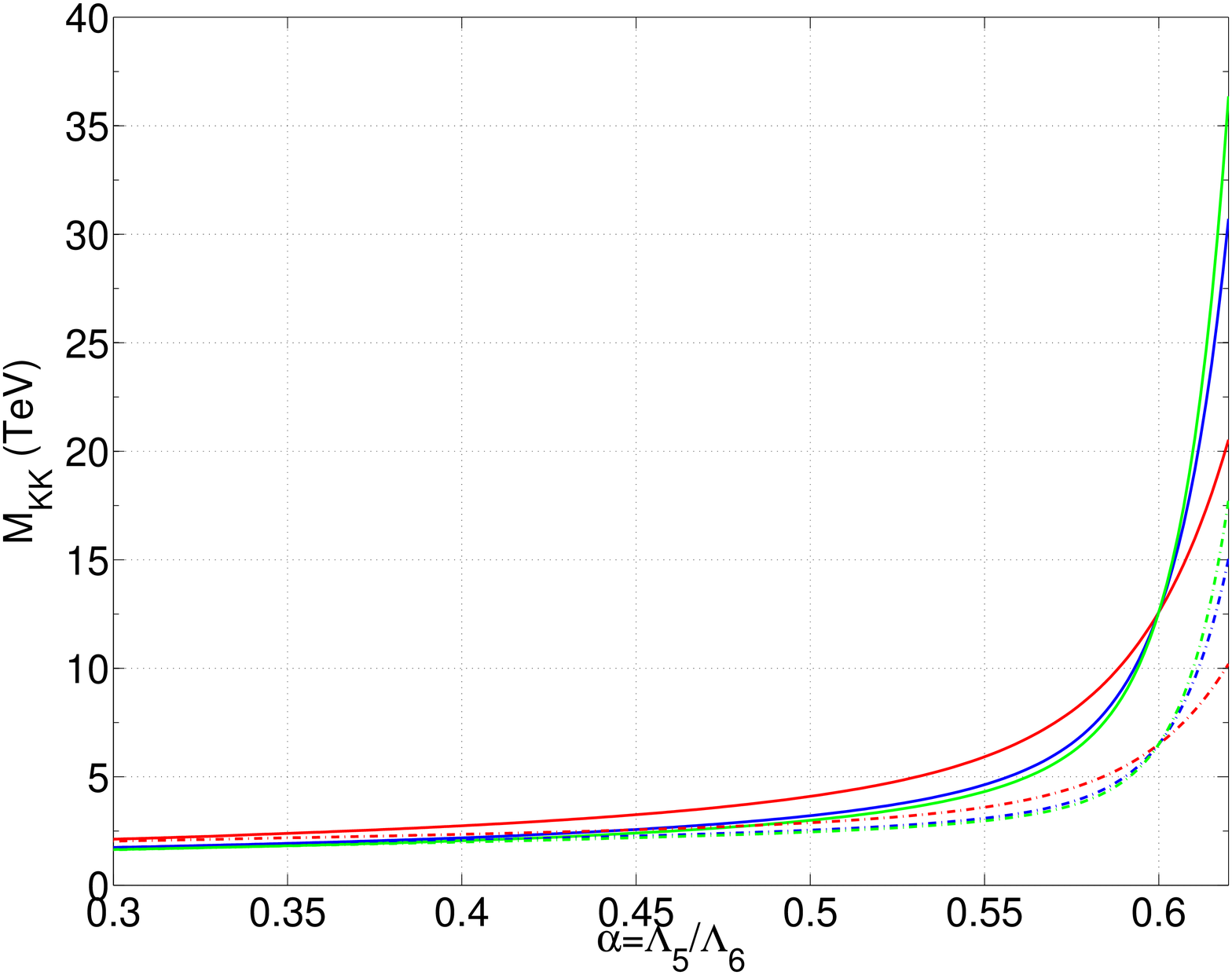}&
\includegraphics[width=2.8in]{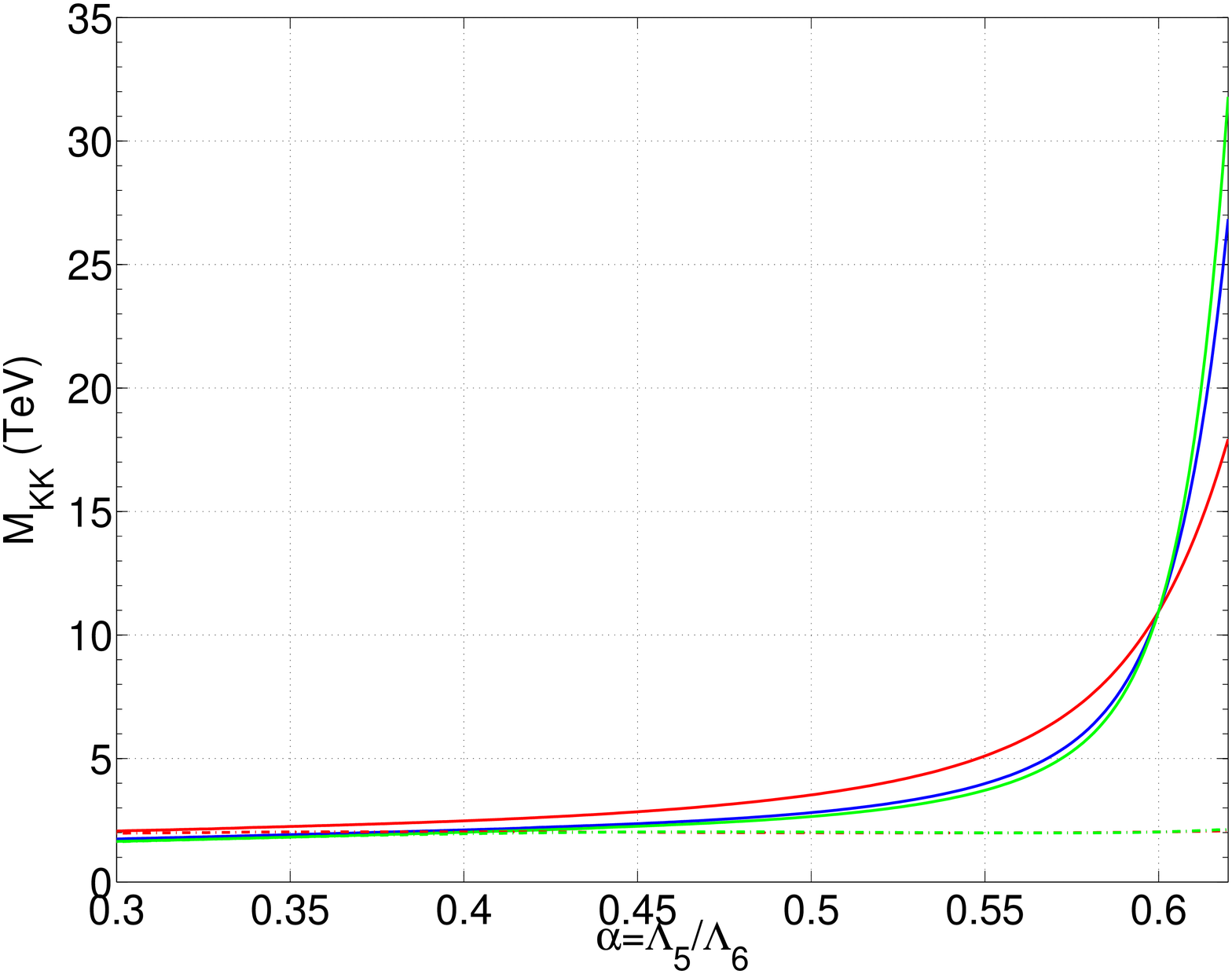}\\
\end{tabular}
\caption{\footnotesize The lower bound on $M_{KK}\equiv\frac{k}{\Omega}$ arising from the EW observable $s_Z^2$. With a bulk $SU(2)\times U(1)$ gauge symmetry (top) and a bulk $SU(2)_R\times SU(2)_L\times U(1)$ custodial symmetry (bottom). Here the fermions are localised to the IR brane (solid line) or the UV brane (dot-dash line) while the gauge fields propagate in 6D (red), 8D (blue) and 10D (green). The Higgs is localised w.r.t $r$ such that $\Omega=10^{15}$.\normalsize }
\label{SZcons}
\end{center}
\end{figure}

Moving on to consider the relative gauge couplings, plotted in figure \ref{couplingfig }, one can see the combined effects of warping and volume enhancement. Due to warping the KK modes will be localised towards the IR brane, while the zero mode is flat. So when the additional dimensions are warped ($J\neq 0$), the volume of those extra dimensions will scale the KK modes very differently to that of the zero mode. The result is a significant enhancement or suppression of the relative gauge couplings. The effect of this on the EW constraints is plotted in figure \ref{SZcons}. The constraints for $\alpha>\frac{3}{5}$ ($J>0$) are not plotted since the gauge fields would become strongly coupled and hence the tree level analysis would not be valid. Although one could assume the constraints would be large. There also appears to be an overall lower bound, corresponding to $M_{KK}\gtrsim 2$ TeV, that arises from the difficulty in getting $F_n^2$ or $F_nF_\psi < 1$. 
\section{Conclusion}
The model presented in the second half of these proceedings can only really be considered as a toy model, whose sole purpose was to demonstrate the phenomenological implications of having additional ($D>5$) warped dimensions. Having said that, the volume scaling of the relative coupling as well as the additional KK modes, are generic effects that appear in all spaces looked at, including solutions from string theory such as \cite{Klebanov:2000hb}\cite{Archer:2010hh}. If one is to speculate that all strongly coupled Yang Mills theories have an AdS dual space then restricting ones phenomenological studies to that of $\rm{AdS}_5$ may lead to misinterpretation of signals at the LHC. Here it has been demonstrated that the EW constraints, on models with warped extra dimensions, can be reduced with or without a custodial symmetry or UV localised fermions, by the inclusion of an additional dimension warped in the opposite direction to that of the four large dimensions. It has also been demonstrated that gauge fields propagating in spaces with additional warped dimensions would have a very different KK spectrum, couplings and hence phenomenology to that of the RS model.  \newline     

\bibliographystyle{iopart-num}
\bibliography{Bibliography}

\end{document}